\documentclass[aps,prd,preprintnumbers,nofootinbib,twocolumn]{revtex4}
\usepackage{bm}
\usepackage{latexsym}
\usepackage{dcolumn}
\usepackage{amsmath,amsfonts,amssymb}
\usepackage{amsthm}

\def\be {\begin{equation}}
\def\ee {\end{equation}}
\def\bea {\begin{eqnarray}}
\def\eea {\end{eqnarray}}
\def\bc {\begin{center}}
\def\ec {\end{center}}
\def\bfg {\begin{figure}}
\def\efg {\end{figure}}
\def\bi {\begin{itemize}}
\def\ei {\end{itemize}}

\def\beq{\begin{equation}}
\def\eeq{\end{equation}}
\def\br{\begin{eqnarray}}
\def\er{\end{eqnarray}}
\newcommand{\eel}[1] {\label{#1}\end{equation}}

\newcommand{\bdm}{\begin{displaymath}}
\newcommand{\edm}{\end{displaymath}}

\begin{document}
\title{Discreteness of Space from    Anisotropic Spin-Orbit Interaction}

\author{Ahmed Farag Ali $^1$} \email[email: ]{ahmed.ali@fsc.bu.edu.eg}
\author{Barun Majumder $^2$} \email[email: ]{barunbasanta@gmail.com}

\affiliation{$^1$Dept. of Physics, Faculty of Sciences, Benha University, Benha 13518, Egypt. \\}

\affiliation{$^2$ University of Tennessee,
Knoxville, USA
}

\begin{abstract}
Various approaches to Quantum  Gravity  suggest an existence of a minimal measurable length. The cost to have such  minimal length could be modified  uncertainty principle, modified dispersion relation, non-commutative geometry or breaking of continuous Lorentz symmetry. In this paper, we propose that minimal length can be obtained naturally through spin-orbit interaction. We consider Dresselhaus anisotropic spin-orbit interaction as the perturbative Hamiltonian. When applied to a particle, it implies that the space, which seizes this particle, should be quantized in terms of units that depend on particle's mass. This suggests that all measurable lengths in the space are quantized in units depending on existent mass and the Dresselhaus coupling constant. On one side, this indicates a breakdown of the space continuum picture near the scale of tabletop experiments, and on the other side, it proposes that spin-orbit interaction is a possible quantum gravity effect at low energy scale that leads naturally to space quantization.
\end{abstract}

\maketitle

Physicists have developed various approaches to quantum gravity, however still today we do not have a conclusive evidence of quantum gravitational effects from experiments. The quantization of space from various theories predicts the existence of minimal length and its manifestation is widely predicted to exist in the form of a generalized uncertainty principle \cite{a1,a2,a3,a4,b1}. A more detailed review on the topic can be found in \cite{arev}. The effects of quantum gravity have been investigated in various experiments and observations. For example, observations from gamma-ray bursts are still an open avenue for the investigations \cite{test1,test2} but far from any conclusive evidence \cite{test3,test4}. Another promising path is to perform quantum mechanical tabletop experiments to look for deviations from standard quantum mechanical results as a consequence of modifying the Heisenberg uncertainty principle based on the predictions from a theory of quantum gravity \cite{a4,test5}. Recently there has been a suggestion that experiments with a quantum optical oscillatory system can provide a measurement of the canonical commutator by optical interferometric techniques. The set up is highly effective in investigating the bounds on quantum gravitational parameters in the generalized uncertainty principle \cite{test6}. Bekenstein \cite{test7,test8} proposed a tabletop experiment arguing that the present ultrahigh vacuum and cryogenic technology may have enough sensitivity to detect signatures of quantum gravity. The main objective of his table top experiment was to prove that at length scales of the order of Planck length the structure of space is not a smooth manifold. In other words, insights on the discreteness of space at Planck scale may be drawn from state-of-art tabletop experimental setups.
\par
If we only consider elementary particles as probes for quantum gravity we would need to accelerate the particles which corresponds to very high energies which is not feasible even within the realm of future particle accelerators. As an alternative approach to investigate the nature of quantum gravity we must try to relate our understanding to the present quantum laboratory experiments. In this paper we will discuss a spin-orbit coupling perturbative Hamiltonian that is reponsible for splitting of energy bands in condense matter systems with broken inversion symmetry commonly known as the Dresselhaus effect \cite{d1}. Later we show that if we consider such an Hamiltonian as a perturbation (whose effect has already been observed in condensed matter systems) and quantize the system with respect to Schr$\ddot{o}$dinger equation, we essentially arrive at the \emph{discreteness of space}. A recent study has shown that a similar spin-orbit coupling whose strength is of the same order as that of Dresselhaus coupling can generate a Lorentz violation in a non-commutative theory \cite{epl1}.
\par
Spin-orbit coupling is an artifact of special theory of relativity where the electric and magnetic fields of a moving electron in its own reference frame can undergo transformations and interact with it's own spin to break spin state degeneracies. Rashba \cite{d2} and Dresselhaus \cite{d1} spin-orbit coupling are very important in the field of spintronics which in terms of strength are similar and they are realized in systems with broken inversion symmetry. In semiconductor quantum well systems it has been found that if the strengths of the Rashba and Dresselhaus interactions are equal then SU(2) symmetry is realized, however the SU(2) is broken in the presence of a cubic Dresselhaus coupling \cite{well1,well2}.
\par
Let us consider the minimal Hamiltonian ($H_{RD}$) with Rashba and Dresselhaus coupling up to third-order in momentum with all the symmetry allowed terms in the $p_z =0$ plane \cite{m1}
\begin{equation}
H_{RD} = H_0 + \frac{1}{\hbar^2} \alpha(p) p_y \sigma_z  ~~,
\label{rdeq1}
\end{equation}
where $p_y$, $p_z$ are the cartesian components of the momentum and ${\bf \sigma}$ are the Pauli matrices for the spin degree of freedom. The Hamiltonian can be derived from a ${\bf k\cdot p}$ perturbation theory. The free particle parabolic part of the Hamiltonian is
\begin{equation}
H_0 = \frac{p_y^2}{2m_y^*} ~~,
\end{equation}
where $m_y^*$ is the effective mass along orthogonal direction $p_y$. The pure linear and cubic Rashba-Dresselhaus limit is recovered when
\begin{equation}
\alpha (p) = \alpha_l + \frac{\alpha_c}{\hbar^2} p_y^2 ~~,
\end{equation}
where $\alpha_l$ is the linear spin-orbit coupling constant and $\alpha_c$ account for p-cubic anisotropic interactions. The experimental values of $\alpha_l$ and $\alpha_c$ can be found in TABLE I of \cite{m1}. For a purely cubic Dresselhaus spin-polarization pattern let us consider only the anisotropic coupling term ($\alpha_l =0$). The Hamiltonian of Eq.(\ref{rdeq1}) can be written as
\begin{equation}
H_{RD} = \frac{p_y^2}{2m_y^*} + \frac{\alpha_c}{\hbar^4}p_y^3 \sigma_z ~~.
\label{rdeq2}
\end{equation}
From the study in \cite{b1} we know that a cubic order term in the Hamiltonian is crucial for achieving discreteness of space however we need to consider  corrections to the uncertainty principle. In this present study we consider an interaction found naturally in many condense matter systems with a cubic term in momentum to get discreteness of space without any other assumptions. Let us study a quantum particle in a box problem with such an Hamiltonian of Eq.(\ref{rdeq2}) considering the cubic Dresselhaus anisotropic term as a perturbation. With the diagonal Pauli matrix $\sigma_z$ we have to solve a Schr$\ddot{o}$dinger equation of the form
\begin{equation}
\left[-\frac{\hbar^2}{2m_p}\frac{\partial^2}{\partial x^2} + \frac{i\alpha_c}{\hbar}\frac{\partial^3}{\partial x^3}\right] \psi = E \psi ~~,
\label{scheq1}
\end{equation}
from $x=0$ to $L$ with boundary conditions $\psi(0)=\psi(L)=0$. A general solution to the equation in leading order in $\alpha_c$ can be written as
\begin{equation}
\psi = A e^{ik(1+k\alpha_c m_p/\hbar^3)x} + B e^{-ik(1-k\alpha_c m_p/\hbar^3)x} + C e^{\frac{-ix\hbar^3}{2\alpha_c m_p}} ~~,
\end{equation}
where $k=\sqrt{2m_pE/\hbar^2}$ and $A$, $B$ and $C$ are arbitrary integration constants to be fixed by boundary conditions. With $\alpha_c=0$ we must get the energy quantization in the form $E_n=\frac{n^2 \pi^2 \hbar^2}{2m_pL^2}$.  This requires for $\alpha_c \rightarrow 0$, $\vert C\vert \rightarrow 0$. The boundary condition $\psi(0)=0$ gives the condition $B=-A-C$. This along with the condition $\psi(L)=0$ gives an equation
\begin{align}
2A\sin (kL) \sin \left(\frac{\alpha_c m_p k^2 L}{\hbar^3}\right) &= 2iA\sin (kL) \cos \left(\frac{\alpha_c m_p k^2 L}{\hbar^3}\right) \nonumber \\
&- C e^{-ikL(1-k\alpha_c m_p/\hbar^3)} \nonumber \\
&+ C e^{-iL\hbar^3/2\alpha_c m_p}
\end{align}
If we separate the real and imaginary parts then we get the following equations:
\begin{align}
2A\sin (kL) \sin \left(\frac{\alpha_c m_p k^2 L}{\hbar^3}\right) &+C \cos \left(kL(1-k\alpha_c m_p/\hbar^3)\right) \nonumber \\
&= C \cos \left(\frac{L\hbar^3}{2\alpha_c m_p}\right)
\label{cond1}
\end{align}
and
\begin{align}
2A\sin (kL) \cos \left(\frac{\alpha_c m_p k^2 L}{\hbar^3}\right) &+C \sin \left(kL(1-k\alpha_c m_p/\hbar^3)\right) \nonumber \\
&= C \sin \left(\frac{L\hbar^3}{2\alpha_c m_p}\right)
\label{cond2}
\end{align}
As mentioned earlier when $\alpha_c \rightarrow 0$, $\vert C \vert \rightarrow 0$ which is the usual energy quantization for a quantum particle in a box and we get the condition $kL=q\pi ~(q\in \mathbb{N})$. But when $\alpha_c \neq 0$ (its experimental value can be found in \cite{m1}), $kL=q\pi$ must still hold. As $\vert C \vert \neq 0$ when $\alpha_c \neq 0$ the following additional relation must be satisfied to satisfy the boundary conditions given by Eqns.(\ref{cond1}) and (\ref{cond2}):
\begin{equation}
L= q\pi \frac{2\alpha_c m_p}{\hbar^3} + {\cal O}(\alpha_c^2) ~~~~~\text{where}~~~~~ q \in \mathbb{N}.
\label{quan1}
\end{equation}
So in order to contain a particle in the box, length must be quantized according to Eq.(\ref{quan1}). Observe that the length is quantized in terms of the mass of the particle ($m_p$). This indicates a dynamical relation between particle's mass and space quantization. If the space has a discrete nature, its  discreteness unit will depend on the mass contained in this space. The discreteness of space in that sense behaves like curvature of space. Both are varying with mass. Similar results of length quantization have been obtained with the generalized uncertainty principle \cite{b1} and are also found in Loop Quantum Gravity \cite{lqg1}. However in those approaches the length quantization comes in units of Plank length ($l_{PL}$). Here we show that a perturbative spin-orbit interaction with broken SU(2) symmetry can automatically give discreteness of space in terms of the mass of the particle without the need of any extra assumptions. This indicates that Spin-Orbit interaction which is a natural interaction in many condense matter systems is a possible  quantum gravity effect. Even though studied in a different context, our result is analogous to the area quantization which is evident in loop quantum gravity where $\Delta A\propto M^2$, where $A$ is the area of event horizon and $M$ is the black hole mass \cite{hod}. Also the mass of a Schwarzschild black hole can be shown to be quantized \cite{ma1,ma2,ma3} which agrees with Bekenstein's argument \cite{bek1}. The constant associated in loop quantum gravity is the Immirzi parameter \cite{ash1} and is our case a natural anisotropic coupling $\alpha_c$. More careful investigation is needed on this point. It will be also interesting to investigate such discreteness in two and three dimensions to look at the discreteness of area and volume. It will be interesting to study the phenomenological implications of such discreteness of space in tabletop experiments. We hope to report on these in the future.

%
%


\begin{thebibliography}{100}

\bibitem{a1} D. Amati, M. Ciafaloni, G. Veneziano, Phys. Lett. B {\bf 216} (1989) 41.
\bibitem{a2} M. Maggiore, Phys. Lett. B {\bf 304} (1993) 65.
\bibitem{a3} J. Magueijo, L. Smolin, Phys. Rev. Lett. {\bf 88} (2002) 190403.
\bibitem{a4} S. Das, E. C. Vagenas, Phys. Rev. Lett. {\bf 101} (2008) 221301.

\bibitem{b1} A. F. Ali, S. Das and E. C. Vagenas, Phys.\ Lett.\  B {\bf 678} (2009) 497.

\bibitem{arev} S. Hossenfelder, Living Rev. Relativity {\bf 16} (2013) 2.


\bibitem{test1} Amelino-Camelia, G., Ellis, J., Mavromatos, N. E., Nanopoulos, D. V. and
Sarkar, S., Nature {\bf 393} (1998) 763-765.
\bibitem{test2} Jacob, U. and Piran, T., Nature Phys. {\bf 7} (2007) 87-90.
\bibitem{test3} Abdo, A. A. et al., Nature {\bf 462} (2009) 331-334.
\bibitem{test4} Tamburini, F., Cuofano, C., Della Valle, M. and Gilmozzi, R., Astron. Astrophys. {\bf 533} A71 (2011).
\bibitem{test5} Ali, A. F., Das, S. and Vagenas, E. C., Phys. Rev. D {\bf 84} (2011) 044013.

\bibitem{test6} Pikovski, I., Vanner, M. R., Aspelmeyer, M., Kim, M.S. and Brukner, C., Nature Phys. {\bf 8} (2012) 393-397.

\bibitem{test7} J. D. Bekenstein, Phys. Rev. D {\bf 86} (2012) 124040.
\bibitem{test8} J. D. Bekenstein, Found Phys {\bf 44} (2014) 452-462.

\bibitem{d1} G. Dresselhaus, Phys. Rev. {\bf 100} (1955) 580.

\bibitem{epl1} S. Aghababaei and G. Rezaei, Euro. Phys. Lett. {\bf 132} (2020) 11002.

\bibitem{d2} E. I. Rashba, Sov. Phys. Solid State {\bf 2} (1960) 1109.

\bibitem{well1} J. D. Koralek {\it et. al.}, Nature {\bf 458} (2009) 610-613.
\bibitem{well2} Lorenz Meier {\it et. al.}, Nature Physics {\bf 3} (2007) 650-654.

\bibitem{m1} C. Autieri, P. Barone, J. Slawinska and S. Picozzi, Phys. Rev. Materials {\bf 3} (2019) 084416.

\bibitem{lqg1} T. Thiemann, J. Math. Phys. {\bf 39} (1998) 3372.

\bibitem{hod} S. Hod, Phys. Rev. Lett. {\bf 81} (1998), 4293; S. Hod, Gen. Rel. Grav. {\bf 31} (1999), 1639.

\bibitem{ma1} J. Makela, P. Repo, Phys. Rev. D {\bf 57} (1998) 4899.
\bibitem{ma2} O. Obregon, M. Sabido, V.I. Tkach, Gen. Relativ. Gravit. {\bf 33} (2001) 913.
\bibitem{ma3} B. Majumder, Phys. Lett. B {\bf 701} (2011) 384.

\bibitem{bek1} J.D. Bekenstein, Phys. Rev. D {\bf 7} (1973) 2333; J.D. Bekenstein, Phys. Rev. D {\bf 9} (1974) 3292;

\bibitem{ash1} A. Ashtekar, J. Baez, A. Corichi and K. Krasnov, Phys. Rev. Lett. {\bf 80} (1998) 904.


\end{thebibliography}
\end{document}